\documentstyle [12pt]{article}
\advance\hoffset by -.5 in
\advance\voffset by -1 in 
\begin{document}

\def\pz{\p_z}
\def\a{\alpha}
\def\b{\beta}
\def\g{\gamma}
\def\d{\delta}
\def\e{\epsilon}
\def\l{\lambda}
\def\x{\xi}            
\def\f{\phi}
\def\j{\psi}
\def\t{\tau}
\def\L{\Lambda}
\def\l{\lambda}
\def\p{\partial}
\def\o{\omega}
\def\O{\Omega}
\def\z{\zeta}
\def\Tau{{\rm T}}
\def\cA{{\cal A}}
\def\tcA{\tilde{\cal A}}
\def\tW{{\bf\tilde W}}
\def\cK{{\cal K}}
\def\tf{\tilde{f}}
\def\tg{\tilde{g}}
\def\th{\tilde{h}}
\def\bL{{\bf L}}
\def\bM{{\bf M}}
\def\bN{{\bf N}}
\def\bP{{\bf P}}
\def\bW{{\bf W}}
\def\n{N}
\def\rarr{\rightarrow}
\def\mp{\mapsto}
\def\hL{\hat{L}}
\def\hw{\hat w}
\def\res{{\rm res\:}}
\def\tr{{\rm tr}~}
\def\tres{\tr\res}
\def\cF{{\cal F}}
\def\et{\eta}
\def\cV{{\cal V}}
\def\s{\sum}
\def\cL{{\cal L}}
\def\cH{{\cal H}}
\def\hw{\hat{w}}
\def\bg{{\bf g}}
\def\un{\underline}
\def\({\left(}
\def\){\right)}
\def\Lm{\Lambda}
\hfill stab2.tex \today\\

\vspace{.1in}
\centerline{{\bf Chains of KP, Semi-infinite 1-Toda Lattice Hierarchy}}
\centerline{{\bf and Kontsevich Integral}}

\vspace{.2in}

\centerline{{\bf L. A. Dickey}}

\vspace{.1in} 
\centerline{University of Oklahoma\footnote{e-mail: ldickey@math.ou.edu}, 
Norman, OK}

\begin{abstract}

There are well-known constructions of integrable systems 
which are chains of infinitely many copies of the equations of the KP hierarchy
``glued'' together with some additional variables, e.g., the modified KP hierarchy.
Another interpretation of the latter, in terms of infinite 
matrices, is called the 1-Toda lattice hierarchy. One way infinite reduction of
this hierarchy has all solutions in the form of sequences of expanding
Wronskians. We define another chain of the KP equations, also with solutions
of the Wronsksian type, which is characterized by the property to
stabilize with respect to a gradation. Under some constraints imposed,
the tau functions of the chain are the tau functions associated with the
Kontsevich integrals.
\end{abstract}

\vspace{.3in}
{\bf 0. Introduction.}\\

This paper was motivated by the following arguments. There are well-known chains
of infinitely many copies of the equations of the KP hierarchy ``glued''
together with some variables, like, e.g., modified KP (see Eqs(1a,b,c)
below). The latter is a sequence of dressing operators of the KP hierarchy $\{\hw_N\}$
along with ``gluing'' variables $\{u_N\}$. All these variables make a large integrable system.
The chain (1a,b,c) has another interpretation, 
in terms of infinite matrices, which is called the 1-Toda lattice hierarchy (see
[1-3]). There exist different reductions of this chain, e.g., modified
KdV, or another reduction, a semi-infinite chain for which all $\hw_N$ with 
negative $N$ are trivial, $\hw_N=1$, and $\hw_N$ with positive $N$ are 
$P_N\p^{-N}$ where $P_N$ is an $N$th order differential operator (the
corresponding matrices of the 1-Toda lattice hierarchy also are semi-infinite). 
It can be shown (see below)
that all the solutions are sequences of well-known Wronskian solutions
to KP, each $\hw_N$ being represented by a determinant of $N$th order. Every
next determinant is obtained from the preceding one by an extension of the 
Wronskian when a new function is added to the existing ones. 

There is another situation where one deals with a sequence
of Wronskian solutions of increasing order. This time the Wronskians are
not obtained by a successive extension. The rule is more complicated. We talk
about the so-called Kontsevich integral [4-6] which has its origin in 
quantum physics. This is an integral over the group U(N) which is a function
of a matrix, invariant with respect to the matrix conjugation, i.e., a function 
of eigenvalues $\l_i$ of the matrix. The main fact about 
the Kontsevich integral is that it is a tau function of the KP hierarchy
of the Wronskian type in variables $t_i=\sum_k\l_k^{-i}$. The dimension
of the Wronskian is $N$. The sequence of Wronskians has, in a sense, 
a limit when $N\rarr\infty$. More precisely, this is a
stable limit. There is some grading and the terms of a fixed weight
stabilize when $N\rarr\infty$: they become independent of $N$ when $N$
is large enough. The stable limit belongs to the $n$th reduction of KP
($n$th GD) and, besides, satisfies the string equation.

The question we try to answer here is whether the sequence of Kontsevich tau
functions is interesting by itself, not only by its limit. Is it possible
to complete it with ``gluing'' variables to obtain a chain of related
KP equations similar to (1a,b,c)? The answer is positive (see Sect. 2.1,
Eqs(a,b,c,d)). Unfortunately, we do not know a matrix version of this chain 
like that of the 1-Toda lattice hierarchy.

Thus, in this paper we define the ``stabilizing'' chain of KP, study
its solutions and demonstrate that they are exactly those which are represented
by the Kontsevich integral. In Appendix we briefly, skipping all the
calculations, show the way from the Wronskian solutions to the Kontsevich
integral. This is actually the conversion of Itzykson and Zuber's [5]
reasoning, and the reader can find the skipped detail there\footnote{I
have a complete version of this Appendix with all the detail.
If someone is interested, I can send that to him/her.}

I am thankful to H. Aratyn discussions with whom were very helpful.\\

{\bf 1. Preliminaries. Semi-infinite 1-Toda lattice hierarchy.}\\

{\bf 1.1.} Recall some basic facts about the modified KP and 1-Toda lattice
hierarchy (see [3]). The modified KP hierarchy is a collection of the following
objects ($\p_k=\p/\p t_k,~\p=\p_1$): 
$$\hw_\n(\p)=1+w_{\n1}\p^{-1}+w_{\n2}\p^{-2}+...~~{\rm where}~\n\in
{\bf Z}$$ and $\{u_\n\}$, $\n\in {\bf Z}$ and relations:
$$(\p+u_\n)\hw_\n=\hw_{\n+1}\p, \eqno{(1.a)}$$
$$\p_k\hw_\n=-(L_\n^k)_-\hw_\n,~{\rm where}~L_\n=\hw_\n\p\hw_\n^{-1}, \eqno{(1.b)}$$
$$\p_ku_\n=(L_{\n+1}^k)_+(\p+u_\n)-(\p+u_\n)(L_{\n}^{k})_+.\eqno{(1.c)}$$
Notice that multiplying Eq.(1.c) by $(\p+u_\n)^{-1}$ on the right
and taking the residue we get an equivalent form of this equation:
$$\p_ku_\n=-\res(\p+u_\n)(L_\n^k)_+(\p+u_\n)^{-1}. \eqno{(1.c')}$$
Let us construct a both way infinite matrix $\bW$ with elements
$$W_{ij}=\left\{\begin{array}{cc}w_{i,i-j},&{\rm when}~j\leq i\\0&{\rm
otherwise}\\\end{array}\right.$$ Then a proposition holds:\\

{\bf Proposition} (see [3]). {\sl Operators $\{\hw_\n\}$ satisfy the modified KP,
along with some $\{u_\n\}$, if and only if the matrix $\bW$ satisfies the
discrete KP (1-Toda lattice hierarchy):$$\p_k\bW=-(\bL^k)_-\bW$$ where
$\bL=\bW\L\bW^{-1}$, $\L$ is the matrix of the shift: $(\L)_{ij}=\d_{i,j-1}$
and the subscript ``$-$'' symbolizes the strictly lower triangular part of a
matrix}.\\

Suppose we have a special solution such that $\hw_\n=1$ and $u_{\n-1}=0$ when 
$\n\leq 0$. Let $P_\n=\hw_\n\p^\n$. Then $P_0=1$, Eq.(1.a) implies $(\p+u_\n)P_\n=P_
{\n+1}$ and therefore $P_\n$ is an $\n$th order differential monic operator when 
$\n>0$. Then $\hw_\n=1+w_{\n1}\p^{-1}+...+w_{\n\n}\p^{-\n}$. The matrix
$\bW$ is a direct sum of two semi-infinite blocks, one is the unity and
one is $(W_{ij})$ with $i$ and $j$ $\geq 0$. We shall show that
all the solutions are simply the well-known Wronskian solutions of KP.\\

{\bf 1.2.}  {\bf Lemma.} {\sl Let $\hw_N$ be a solution of the semi-infinite 
chain of equations (1.a-c) with $\n=0,1,2,...$ and $P_N=\hw_N\p^N$. There 
exists a sequence of linearly independent functions
$y_0,y_1,...$ such that $P_\n y_i=0$ when $i=0,...,\n-1$ and $(\p^k-\p_k)y_i=0$
for all $k$ and $i$.}\\ 

{\em Proof.} Suppose, $y_i$'s are already constructed for $i<\n-1$ (if $\n=1$,
nothing is supposed). Then $P_{\n-1}y_i=0$ and $P_\n y_i=0$ for $i=0,...,\n-
2$ since $(\p+u_{\n-1})P_{\n-1}=P_\n$. The kernel of the operator $P_\n$ is
$\n$-dimensional, therefore there is one more function $y_{\n-1}$ independent
of $y_0,...,y_{\n-2}$ such that $P_\n y_{\n-1}=0$. 

First of all, let us prove that if a function, in this case $y_{N-1}$
but this is a general fact, belongs to the kernel of $P_N$ then so does
$(\p^k-\p_k)y_{N-1}$. We have $$0=\p_k(P_\n y_{\n-1})=
(\p_kP_\n)y_{\n-1}+P_\n(\p_ky_{\n-1})=-(\hw_\n\p^k\hw_\n^{-1})_-\hw_\n\p^\n y_{\n-1}
+P_\n(\p_ky_\n)
$$ $$=-\hw_\n\p^\n\p^ky_{\n-1}+(L_\n^k)_+P_\n y_{\n-1}+P_\n (\p_ky_{\n-1})$$ 
The middle term
vanishes since $P_\n y_{\n-1}=0$ and $(L_\n^k)_+$ is a differential operator.
Thus, $P_\n(\p^k-\p_k)y_{\n-1}=0$, and $(\p^k-\p_k)y_{\n-1}$ is in Ker $P_N$. Now,
$$(\p^k-\p_k)y_{\n-1}=\sum_{i=0}^{\n-1} 
A_{ki}y_i.$$ The coefficients $A_{ki}$ do not depend on $t_1$. We have
$$(\p^l-\p_l)(\p^k-\p_k)y_{\n-1}=A_{k,\n-1}\sum_
{i=0}^{\n-1}A_{li}y_i-\sum_{i=0}^{N-1}(\p_lA_{ki})y_i,$$
$$(\p^k-\p_k)(\p^l-\p_l)y_{\n-1}=A_{l,\n-1}\sum_
{i=0}^{\n-1}A_{ki}y_i-\sum_{i=0}^{N-1}(\p_kA_{li})y_i.$$
Operators $(\p^k-\p_k)$ and $(\p^l-\p_l)$ commute, hence
$$\sum_{i=0}^{N-1}(\p_l+A_{l,N-1})A_{ki}\cdot y_i=
\sum_{i=0}^{N-1}(\p_k+A_{k,N-1})A_{li}\cdot y_i$$ and, by virtue of the
linear independence of $y_i$ as functions of $t_1$, $$(\p_l+A_{l,N-1})A_{ki}
=(\p_k+A_{k,N-1})A_{li}.\eqno{(1.2.1)}$$ This is the compatibility condition  
of the equations $$(\p_k+A_{k,N-1})m_i=A_{ki}.$$ We can find $m_i$, with 
$m_{N-1}=1$,
and then $$(\p^k-\p_k)y_{\n-1}=\sum_{i=0}^{\n-1}(\p_k+A_{k,N-1})m_i\cdot y_i.$$
Let $\tilde y_{N-1}=\sum_0^{N-1}m_iy_i$. It is easy to see that $$(\p^k-
\p_k)\tilde y_{N-1}=A_{k,N-1}\tilde y_{N-1}.$$ Now, if we solve the
system $$\p_k\phi_{N-1}=\phi_{N-1}A_{k,N-1}$$ consistent by virtue of
(1.2.1) where $i=N-1$ then $y^*_{N-1}=\tilde y_{N-1}\phi_{N-1}$ will 
satisfy the equation $$(\p^k-\p_k)y^*_{N-1}=0,$$ q.e.d.\\

{\bf 1.3. Proposition.} {\sl The general construction of solutions to the 
semi-infinite 1-Toda hierarchy, or, equivalently, the chain $(1.a,b,c)$ is the 
following: let $\{y_i\}$ be a sequence of functions of variables $t_1=x,
t_2,...$ having the property $(\p^k-\p_k)y_i=0$. Then
$$\hw_\n={1\over W_\n}\left|\begin{array}{cccc}y_{0}&...&y_{\n-1}&1\\
y'_{0}&...&y'_{\n-1}&\p\\ .&.&.&.\\y^{(\n)}_{0}&...&y^{(\n)}_
{\n-1}&\p^\n\end{array}\right|\cdot\p^{-\n}~~{\rm where}~W_\n=W(y_0,...,y_{\n-1})~
{\rm is~the~Wronskian}\eqno{(1.3.1)}$$ and $u_\n=-\p\ln (W_{\n+1}/W_\n)$.}\\

{\em Proof.} In one way, the proposition is almost proved by the
preceding analysis. It only remains to notice that $P_\n=\hw_\n\p^\n$ where $\hw_\n$ is
given by Eq.(1.3.1) is the monic differential operator with the kernel
spanned by $y_0,...,y_{\n-1}$, that $(\p+u_\n)P_{\n}y_{\n}=0$ and
$P_{\n}y_{\n}=W_{\n+1}/W_\n$.

The converse follows from the fact that $\hw_\n$ given by (1.3.1), as it
is well known, is a dressing operator of the KP hierarchy, the operators
$(\p+u_\n)P_\n$ where $u_\n=-\ln(W_{\n+1}/W_\n)$ and $P_{\n+1}$ have the same
kernel spanned by $y_0,...,y_{\n}$ and, therefore, coincide.\\

{\bf 1.4.} The property $(\p^k-\p_k)y_i=0$ implies $$y_i(t_1-{1\over z},
t_2-{1\over 2z^2},t_3-{1\over 3z^3},...)=(1-{1\over z}\p)y_i(t_1,t_2,t_3,
...)\eqno{(1.4.1)}$$ or $y_i(t-[z^{-1}])=(1-\p/z)y_i(t)$, using the common
notation. Indeed,
$$\exp(\sum_1^\infty -(kz^k)^{-1}\p_k)y_i=(\exp\sum_1^\infty -(kz^k)^
{-1}\p^k)y_i=\exp\ln(1-{1\over z}\p)y_i=(1-{1\over z}\p)y_i.$$
The Baker, or wave function is $\hw(\p)\exp\sum_1^\infty t_k\x^k=
\hw(z)\exp\sum_1^\infty t_kz^k$ where
$$\hw_\n(z)={1\over W_\n}\left|\begin{array}{cccc}y_{0}&...&y_{\n-1}&z^{-\n}\\
y'_{0}&...&y'_{\n-1}&z^{-\n+1}\\ .&.&.&.\\y^{(\n)}_{0}&...&y^{(\n)}_
{(\n-1)}&1\end{array}\right|$$ Subtracting the second row divided by $z$
from the first one, then the third divided by $z$ from the second one
etc, we obtain zeros in the last column except the last element which is
1. In the $i$th row there will be elements $y_j^{(i)}-z^{-1}y^{(i+1)}_j
=y_j^{(i)}(t-[z^{-1}])$, according to (1.4.1). Thus, $$\hw_\n(z)=
{W_\n(t-[z^{-1}])\over W_\n(t)}$$ which means that $W_\n(t)$ is the tau
function $\tau_\n(t)$ corresponding to $\hw_\n(z)$, and $u_\n=-
\ln(\tau_{\n+1}/\tau_\n)$.

What are functions $y_i(t)$ with the property $(\p^k-\p_k)y_i=0$?
For example, $\exp\sum_1^\infty t_k\a^k$, linear combinations of
several such functions, and integrals $\int f(\a)\exp(\sum_1^\infty
t_k\a^k)d\s(\a)$. Using the Schur polynomials $p_k(t_1,t_2,...)$
defined by $\exp\sum_1^\infty t_k\a^k=\sum_0^\infty p_k(t)a^k$,
we see that all the above example have a form $y_i(t)=\sum c_{ik}p_k(t)$.
Conversely, any series of this form has the property $(\p^k-\p_k)y_i=0$
since the Schur polynomials have it, as it is easy to see.
Apparently, this is, in a sense, the most general form of such
functions.\\

{\bf 2. Stabilizing chain.}\\

{\bf 2.1.} {\bf Definition}. {\sl The stabilizing chain is a collection of the 
following objects: $$\hw_\n(\p)=1+w_{\n1}\p^{-1}+...+w_{\n\n}\p^{-\n}=P_\n(\p)
\cdot\p^{-\n},~~\n=1,2,3,...$$ and $u_\n,~v_\n,~\n=1,2,3,...$ , and relations:}
$$(\p+u_\n)\hw_\n=(\p+v_{\n+1})\hw_{\n+1}, \eqno{(a)}$$
$$\p_k\hw_\n=-(L_\n^k)_-\hw_\n,~{\rm where}~L_\n=\hw_\n\p\hw_\n^{-1}, \eqno{(b)}$$
$$\p_ku_\n=-\res\left((\p+u_\n)(L_\n^k)_+(\p+u_\n)^{-1}\right), \eqno{(c)}$$
$$\p_kv_{\n+1}=-\res\left((\p+v_{\n+1})(L_{\n+1}^k)_+(\p+v_{\n+1})^{-1}\right). \eqno{(d)}$$

The equations $(c)$ and $(d)$ also can be written as 
$$\p_ku_\n=\res L_\n^k-\res\left((\p+u_\n)L_\n^k(\p+u_\n)^{-1}\right), \eqno{(c')}$$
$$\p_kv_{\n+1}=\res L_{\n+1}^k-\res\left((\p+v_{\n+1})L_{\n+1}^k(\p+v_{\n+1})^{-1}\right) \eqno{(d')}$$
where
$$\res\left((\p+u_\n)L_\n^k(\p+u_\n)^{-
1}\right)=\res\left((\p+v_{\n+1})L_{\n+1}^k(\p+v_{\n+1})^{-1}\right)$$ by
virtue of $(a)$. 

It can happen that starting from some term  $u_\n=v_{\n+1}$ and all
$\hw_\n$ are equal, the chain stabilizes. Then $w_{\n\n}=0$.                                               
  Also it can happen that the chain contains constant
segments and after that again begins to change. These segments can be
just skipped. We shall assume that
all $w_{\n\n}\neq 0$. Nevertheless, some tendency to stabilization
remains, as we shall see below. A grading will be introduced
so that all quantities will be sums of terms of different weights.
We shall see that terms of a given weight stabilize, the
greater is the weight the later the stabilization occurs. The 
stabilization is actually the most important feature of this chain allowing
one to consider the stable limits when $N\rarr\infty$. This is used, e.g., 
in the Kontsevich integral.

The chain is well defined if one proves that 1) the right-hand side of
(b) is a $\Psi$DO of the form $\sum_1^\n a_j\p^{-i}$, 2) vector fields $\p_k$
defined by $(b)$, $(c)$ and $(d)$ respect the relation $(a)$, 3) vector fields
$\p_k$ commute. 

Eq. $(b)$ defines a copy of the KP hierarchy for each $n$. It is clear
that the operator in the r.-h. s. is negative. Rewriting it as
$\p_k\hw_\n=-\hw_\n\p^k+(L_\n^k)_+\hw_\n$, one can see that it contains only
powers $\p^{-i}$ with $i\leq \n$. Thus, 1) is proven.

Now, one has to prove that $$\p_k((\p+u_\n)\hw_\n-(\p+v_{\n+1})\hw_{\n+1})=0$$
if $(a)$ holds. We have
$$\p_k((\p+u_\n)\hw_\n-(\p+v_{\n+1})\hw_{\n+1})$$
$$=\left(\res L_\n^k-\res\left((\p+u_\n)L_\n^k(\p+u_\n)^{-1}\right)\right)\hw_\n$$
$$-\left(\res L_{\n+1}^k-
\res\left((\p+v_{\n+1})L_{\n+1}^k(\p+v_{\n+1})^{-1}\right)\right)\hw_{\n+1}$$
$$-(\p+u_\n)(L_\n^k)_-\hw_\n+(\p+v_{\n+1})(L_{\n+1}^k)_-\hw_{\n+1}.$$
The last two terms are transformed as:
$$-(\p+u_\n)(L_\n^k)_-\hw_\n=-((\p+u_\n)L_\n^k)_-\hw_\n-\res(L_\n^k)\hw_\n$$
$$=-((\p+u_\n)L_\n^k(\p+u_\n)^{-1}(\p+u_\n))_-\hw_\n-\res(L_\n^k)\hw_\n$$
$$=-((\p+u_\n)L_\n^k(\p+u_\n)^{-1})_-(\p+u_\n)\hw_\n$$ $$-\res(L_\n^k)\hw_\n
+\res((\p+u_\n)L_\n^k(\p+u_\n)^{-1})\hw_\n.$$ Similarly,
$$(\p+v_{\n+1})(L_{\n+1}^k)_-\hw_{\n+1}=((\p+v_{\n+1})L_{\n+1}^k(\p+v_{\n+1})^{-1})_-
(\p+v_{\n+1})\hw_{\n+1}$$ $$+\res(L_{\n+1}^k)\hw_{\n+1}
-\res((\p+v_{\n+1})L_{\n+1}^k(\p+v_{\n+1})^{-1})\hw_{\n+1}.$$ Taking into
account $(a)$ and
$$ (\p+v_{\n+1})L_{\n+1}^k(\p+v_{\n+1})^{-1}= (\p+u_\n)L_\n^k(\p+u_\n)^{-1},$$ 
the sum of all the terms is zero. 
 
Before we prove 3), the relations $(c)$ and $(d)$ will be presented in a 
different form. \\

{\bf 2.2.} {\bf Lemma.} {\sl The following remarkable formula holds
$$\left(\p-{w'\over w}\right)^{-1}=\sum_0^\infty\p^{-k-1}{w^{(k)}\over
w}$$ where $w$ is a function.}\\

{\em Proof.} $$\sum_0^\infty\p^{-k-1}{w^{(k)}\over w}\left(\p-{w'\over
w}\right)$$ $$=\sum_0^\infty\p^{-k}{w^{(k)}\over w}-\sum_0^\infty\p^{-k-1}
\left({w^{(k)}\over w}\right)'-\sum_0^\infty\p^{-k-1}{w^{(k)}w'\over
w^2}$$ $$=\sum_0^\infty\p^{-k}{w^{(k)}\over w}-\sum_0^\infty\p^{-k-1}{w^{(k+1)}
\over w}=1.$$

{\bf Corollary.} $$\res\p^k\left(\p-{w'\over w}\right)^{-
1}={w^{(k)}\over w}$${\sl and $$\res P(\p)\cdot \left(\p-{w'\over w}\right)^{-1}
={P(\p)w\over w}$$ for any polynomial $P(\p)$.}\\

Notice that the relation $(a)$ implies $(\p+v_{\n+1})w_{\n+1,\n+1}=0$ where
$(\p+v_{\n+1})w_{\n+1,\n+1}$ is understood as a result of action of the
operator $(\p+v_{\n+1})$ on the function $w_{\n+1,\n+1}$ (not a product!).
Indeed, this is the coefficient in $\p^{-\n-1}$ of the expression in the
r.-h.s. while the l.-h.s. does not contain this term. Thus,
$$v_{\n+1}=-{w'_{\n+1,\n+1}\over w_{\n+1,\n+1}}=-\p\ln w_{\n+1,\n+1}.\eqno{(2.2.1)}$$
Now, $$\p_kv_{\n+1}=-\res\left((\p-w'_{\n+1,\n+1}/ w_{\n+1,\n+1})(L_{\n+1}^k)_+
(\p-w'_{\n+1,\n+1}/ w_{\n+1,\n+1})^{-1}\right)$$
$$=-{(\p-w'_{\n+1,\n+1}/ w_{\n+1,\n+1})(L_{\n+1}^k)_+w_{\n+1,\n+1}\over
w_{\n+1,\n+1}}=-\p{(L_{\n+1}^k)_+w_{\n+1,\n+1}\over w_{\n+1,\n+1}}.\eqno{(2.2.2)}$$
Subtracting $(c')$ and $(d')$ we also have
$$\p_k(u_\n-v_{\n+1})=\res L_\n^k-\res L_{\n+1}^k. \eqno{(2.2.3)}$$
These two equations are equivalent to $(c)$ and $(d)$.

An alternative way to get (2.2.2) is the following. Eq. $(b)$ implies
$\p_k\hw_{\n+1}=-\hw_{\n+1}\p^k+(L_{\n+1}^k)_+\hw_{\n+1}$ and
$$\p_kw_{\n+1,\n+1}=-(L_{\n+1}^k)_+w_{\n+1,\n+1}.$$ Then Eq.(2.2.2) easily
follows from (2.2.1).       \\

{\em Proof of the commutativity of $\{\p_k\}$.} They commute in their
action on all $\hw_\n$ as KP vector fields. Therefore (2.2.1) yields
$\p_k\p_lv_{\n+1}=\p_l\p_kv_{\n+1}$. Finally, $$\p_l\p_k(u_\n-v_{\n+1})
=\res\left([(L_\n^l)_+,L_\n^k]-[(L_{\n+1}^l)_+,L_{\n+1}^k]\right)$$ $$=-\res
\left([(L_\n^l)_-,L_{\n}^k]-[(L_{\n+1}^l)_-,L_{\n+1}^k]\right)$$ $$=-\res
\left([(L_\n^l)_-,(L_{\n}^k)_+]-[(L_{\n+1}^l)_-,(L_{\n+1}^k)_+]\right)$$
$$=-\res\left([L_\n^l,(L_{\n}^k)_+]-[L_{\n+1}^l,(L_{\n+1}^k)_+]\right)$$ 
$$=-\res\left([(L_\n^k)_+,L_{\n}^l]-[(L_{\n+1}^k)_+,L_{\n+1}^l]\right)=
\p_k\p_l(u_\n-v_{\n+1})$$ whence $\p_l\p_ku_\n=\p_k\p_lu_\n$.\\

{\bf 3. Solutions to the chain.}\\

{\bf 3.1.} Let $y_{0\n},...,y_{\n-1\n}$ be a basis of the kernel of the
differential operator $P_\n=\hw_\n\p^\n$: $P_\n y_{i\n}=0$.\\

{\bf Lemma.} {\sl Passing if needed to linear combinations of
$y_{i\n}$ with coefficients depending only on $t_2,t_3,...$, one can always 
achieve} $$ \p_ky_{i\n}=\p^ky_{i\n}\eqno{(i)}$$ and\footnote{The relation (ii) is
what makes this chain different from one considered in the previous
section (modified KP) where it was $y_{i\n}=y_{i,\n+1}$ instead.} $$y_{i\n}=y'_{i,\n+1},~
i=0,...,\n-1.\eqno{(ii)}$$

{\em Proof.} 
Suppose $y_{iN}$, where $i=0,...,\n-1$, are already constructed. 
Let $y_{0,\n+1},...,y_{\n,\n+1}$ be a basis of the kernel
of the operator $P_{\n+1}$: $P_{\n+1}y_{i,\n+1}=0$. The functions
$y'_{0,\n+1},...,y'_{\n,\n+1}$ which belong to the kernel of $(\p+u_\n)P_\n$ 
are linearly independent, otherwise there would be a linear combination of $y_{i,N+1}$
belonging to the kernel of $P_{N+1}$ which is constant (with respect to $t_1=x$)  
while we know that 
$P_{\n+1}1=w_{\n+1,\n+1}\neq 0$ by assumption. Hence, at least one of these 
functions does not belong to ker $P_\n$, let it be $y'_{\n,\n+1}$: $P_\n y'_
{\n,\n+1}\neq 0$. Since all $P_\n y'_{i,\n+1}$ belong to the 1-dimensional kernel 
of $\p+u_\n$, there must be constants $a_i$ such that $P_\n(y'_{i,\n+1}-
a_iy'_{\n,\n+1})=0$. (When we speak about constants, we mean constants with 
respect to $t_1=x$ depending, maybe, on higher times).
Thus $(y_{i,\n+1}-a_iy_{\n,\n+1})'$ form a basis of 
the kernel of $P_\n$. There exist their linear combinations $(y^{(1)}_{i,\n+1})'$
coinciding with $y_{i\n}$: $ (y^{(1)}_{i,\n+1})'=y_{i\n}$ where $i=0,...,\n-1$. This 
yields $\p(\p^k-\p_k)y^{(1)}_{i,\n+1}=0$ and $(\p^k-\p_k)y^{(1)}_{i,\n+1}=
c_{ki}=$const. As in the Lemma 1.2, we can prove that $(\p^k-\p_k)y^{(1)}_{i,\n+1}
\in$Ker $P_{N+1}$, therefore $c_{ki}=0$ since constants do not belong to
the kernel. It remains to consider $y_{\n,\n+1}$.
Since $(\p^k-\p_k)y_{\n,\n+1}=\sum A_iy_{i,\n+1}$, the same reasoning as in
the Lemma of Sec. 1.2. will do the rest.\\

{\bf 3.2. Proposition.} {\sl All solutions to the chain {\it (a-d)} have 
the following structure. Let $y_{i\n}$ where $\n=1,2,...$ and $i=0,...,\n-1$
be arbitrary functions of variables $t_1=x,t_2,...$ satisfying the
relations
$$ \p_ky_{i\n}=\p^ky_{i\n}\eqno{(i)}$$ and
 $$y_{i\n}=y'_{i,\n+1},~
i=0,...,\n-1.\eqno{(ii)}$$ Then
$$\hw_\n={1\over W_\n}\left|\begin{array}{cccc}y_{0\n}&...&y_{\n-1,\n}&1\\
y'_{0\n}&...&y'_{\n-1,\n}&\p\\ .&.&.&.\\y^{(\n)}_{0\n}&...&y^{(\n)}_
{\n-1,\n}&\p^\n\end{array}\right|\cdot\p^{-\n}\eqno{(3.2.1)}$$ where $W_\n=
W_\n(y_{0\n},....y_{\n-1,\n})$
is the Wronskian. Besides,
$$u_\n=-\p\ln{1\over W_\n}\left|\begin{array}{cccc}y_{0\n}&...&y_{\n-1\n}
&y_{\n,\n}\\
y'_{0\n}&...&y'_{\n-1,\n}&y'_{\n,\n}\\ .&.&.&.\\y^{(\n)}_{0\n}&...&y^{(\n)}_
{\n-1,\n}&y^{(\n)}_{\n,\n}\end{array}\right|\eqno{(3.2.2)}$$ where
$y_{\n,\n}=y'_{\n,\n+1}$ by definition and}
$$v_{\n+1}=-\p\ln{1\over W_{\n+1}}\left|\begin{array}{cccc}y_{0\n}&...&y_{\n-1,
\n}&y_{\n,\n}\\
y'_{0\n}&...&y'_{\n-1,\n}&y'_{\n,\n}\\ .&.&.&.\\y^{(\n)}_{0\n}&...&y^{(\n)}_
{\n-1,\n}&y^{(\n)}_{\n,\n}\end{array}\right|.\eqno{(3.2.3)}$$

{\em Proof.} In one way, the proposition follows from the analysis of the 
preceding subsection. Indeed, $P_\n=\hw_\n\p^\n$ given by Eq.(3.2.1) is the
unique differential monic operator having the kernel spanned by
$y_{0\n},...,y_{\n-1\n}$, the latter always can be assumed satisfying the
relations $(i)$ and $(ii)$. Further, $u_\n=-\p\ln P_\n\p y_{\n,\n+1}$
and $v_{\n+1,\n+1}=-\p\ln P_{\n+1}1$ which easily yields Eqs (3.2.2) and
(3.2.3).

Conversely, one must prove that if $\{y_{i\n}\}$ have the properties
$(i)$ and $(ii)$ then Eqs.(3.2.1), (3.2.2) and (3.2.3) present a
solution to the chain $(a-d)$. 

Indeed, KP equations $(b)$ can be obtained in a standard way:
differentiating $\hw_\n\p^\n y_{i\n}=0$ with respect to $t_k$ one gets
$$(\p_k\hw_\n)\p^\n y_{i\n}+\hw_\n\p^\n\p^ky_{i\n}=0$$ or
$$(\p_k\hw_\n)\p^\n y_{i\n}+(L_\n^k)_-\hw_\n\p^\n y_{i\n}=-
(L_\n^k)_+\hw_\n\p^\n y_{i\n}$$ which is zero. The operator $(L_\n^k)_-
\hw_\n\p^\n$ has an 
order less than $\n$. On the other hand, this is a differential operator
since it is equal to $\hw_\n\p^\n-(L_\n^k)_+\hw_\n\p^\n=P_\n-(L_\n^k)_+P_\n$. 
Thus, the differential
operator $(\p_k\hw_\n)\p^\n+(L_\n^k)_-\hw_\n\p^\n$  of order less 
than $\n$ has an $\n$-dimensional kernel and must vanish.

The operators $(\p+u_\n)P_\n\p$ and $(\p+v_{\n+1})P_{\n+1}$ have the same 
kernels spanned by\\ 
$y_{0,\n+1},...,y_{\n,\n+1},1$, therefore they coincide.
We have the equation $(a)$.

From $\p_k\hw_\n=-\hw_\n\p^k+(L_\n^k)_+\hw_\n$ we have $\p_kw_{\n\n}=(L_\n^k)_+
w_{\n\n}$ and $$\p_kv_{\n+1}=-\p[(L_\n^k)_+w_{\n\n}/w_{\n\n}].$$

Finally, applying the operator $\p_k$ to $(a)$ and equating terms of 
zero degree in $\p$, we obtain $\p_ku_\n-\res L_\n^k=\p_kv_{\n+1}-\res L_{\n
+1}^k$. The last two equations are equivalent to $(c)$ and $(d)$
which completes the proof.\\

{\bf 3.3. Solutions in the form of series in Schur polynomials . Stabilization.} \\

Recall that the Schur polynomial are defined by the equation
$$\exp(\sum_1^\infty t_k
z^k)=\sum_0^\infty p_k(t)z^k.$$ A grading can be introduced being prescribed that 
the variable $t_i$ has the weight $i$, the weight of $z$ is $-1$. Then the 
polynomial $p_k(t)$ is of weight $k$. It is easy to verify that the Schur
polynomials have properties $$\p_ip_k=p_{k-i}=\p^ip_k,~~p_k(t)-\z^{-1}
\p p_k(t)=p_k(t_1-\z^{-1}/1,...,t_r-\z^{-r}/r,...)$$ which can be 
obtained from the definition.

Let $$y_{i\n}=\sum_{m=0}^\infty c_m^{(i)}p_{m+N-i},~i=0,...,N-1\eqno{(3.3.1)}$$ 
with some coefficients 
$c_m^{(i)}$ being $c_0^{(i)}=1$. Then the equations (i) and (ii) of
Sect.3.2 hold and Eqs.(3.2.1-3) define solutions for any sets of coefficients
$c_m^{(i)}$. In particular, up to a non-important sign,
$$\tau_N=\left|\begin{array}{lll}y_{0N}^{(N-1)}&......&y_{0N}\\y_{1N}^{(N-
1)}&......&y_{1N}
\\...&......&...\\y_{N-1,N}^{(N-1)}&......&y_{N-1,N}\end{array}\right|.\eqno
{(3.3.2)}$$ Eq.(3.3.1) implies
$$\tau_N=\sum_{m_0,...,m_{N-1}}c_{m_0}^{(0)}...c_{m_{N-1}}^{(N-1)}\left|
\begin{array}{lll}p_{m_0}&......&p_{m_0+N-1}\\p_{m_1-1}&......&p_{m_1+N-2}\\
...&......&...\\p_{m_{N-1}-(N-1)}&......&p_{m_{N-1}}\end{array}\right|\eqno
{(3.3.3)}$$(it is assumed that $p$ with negative
subscripts vanish). \\ 

{\bf Proposition (Itzykson and Zuber).} {\sl Tau functions (3.3.3)  have the 
stabilization property: terms of a weight $l$ do not depend on $N$ when $N>l$.}\\

{\em Proof.}  
The diagonal terms of the determinant are $p_{m_0},p_{m_1},.
..,p_{m_{N-1}}$. All the terms of the determinant are of equal weights, namely, 
$m_0+m_1+...+m_{N-1}=l$. Let us consider a determinant of weight $l$
and prove that all $m_i$ with $i\geq l$ vanish unless the determinant 
vanishes. Suppose that there is some $i\geq l$ such that $m_i\neq 0$. The 
elements of determinant which are located in the $i$th column above 
$p_{m_i}$ have the following
subscripts: $m_0+i,m_1+i-1,...,m_{i-1}+1$. Together with $m_i$, there are 
$i+1$ non-zero integers with a sum $m_0+m_1+...+m_{i-1}+m_i+\sum_{j=1}^ij\leq l
+\sum_{j=1}^i j<i+1+\sum_{j=1}^ij=\sum_{j=1}^{i+1}j$, i.e., less then the sum of
the first $i+1$ integers. This implies that at least two of them coincide. Then
the corresponding rows coincide, and the determinant vanishes.

Thus, if a determinant does not vanish, then, starting from the $l$th row, 
all the diagonal elements are equal $p_0=1$, and all the elements to the left of
the diagonal vanish. The determinant of weight $l$ reduces to a minor of $l$th 
order in the upper left corner, and the terms of weight $l$ are
$$\sum_{m_0+...+m_{l-1}=l}c_{m_0}^{(0)}...c_{m_{l-1}}^{(l-1)}\left|
\begin{array}{lll}p_{m_0}&......&p_{m_0+l-1}\\p_{m_1-1}&......&p_{m_1+l-2}\\
...&......&...\\p_{m_{l-1}-(l-1)}&......&p_{m_{l-1}}\end{array}\right|$$ 
 that does not depend on $N$. \\

{\bf Appendix. From the stabilizing chain to the Kontsevich integral, overview.}\\

{\bf A1.} It is well-known that the so-called Kontsevich integral which
originates in quantum field theory is a tau function of the type (3.3.3) with some
special coefficients $c_m^{(i)}$. We briefly sketch here the way from 
the general solution (3.3.3) to the Kontsevich integral if two additional
requirements are imposed: the stable limit of $\tau_N$ must belong to an $n$th restriction
of KP hierarchy (the $n$th GD) and satisfy the string equation. All the skipped detail of
calculations can be found in the article by Itzykson and Zuber [5] on
which we base our presentation. The only difference is that we do this
in reverse order: not from matrix integrals to tau functions (3.3.3) but
vice versa. It is interesting to see what kind of reasoning and
motivation could lead one from integrable systems to matrix integrals of
the type studied by Kontsevich and to make sure that nothing essential
is lost. Thus, contrary to a tradition, the Kontsevich integral appears
only in the very last lines of the article, {\em v kontse}, which is Russian 
for ``at the end''.

First, we need to make the stable limit of $\tau_N$ belonging to the
$n$th restriction of KP which is equivalent to independence of the $n$th
time, $t_n$.

Usually, when one wishes to make a tau function (3.3.2) independent of
$t_n$, one requires that $\p_ny_{iN}=\a_{iN}y_{iN}$ where $\a_{iN}$ are
some numbers. Then $y_{iN}=\exp(\a_{iN}t_n)y_{iN}(0)$ where $y_{iN}(0)$
does not depend on $t_n$, and $\tau_N=\exp(a_{0N}+...+a_{N_1,N})t_n\cdot
\tau_N(0)$
where $\tau_N(0)$ does not depend on $t_n$. The exponential factor can
be dropped since a tau function is determined up to a multiplication by an
exponential of any linear combination of time variables.

Now, in the problem we are talking about, we do not necessarily wish to
make all $\tau_N$ independent of $t_n$, only their stable limit. Then
it suffices instead of the ``horizontal quasi-periodicity'' $\p_ny_{iN}=
y_{iN}^{(n)}=\a_{in}y_{iN}$ to require the ``vertical periodicity'' 
$\p_ny_{iN}=y_{i+n,N}$. In terms of the series (3.3.1), this means that
$$c_m^{(i+n)}=c_m^{(i)}.\eqno{(A.1.1)}$$ The proof is in [5]. The idea
is clear: when a row which is not one of $n$ last rows is differentiated 
then the resulting determinant has two equal rows. If we consider
only terms of a fixed weight, then they depend only on a minor of a fixed 
size in the upper left corner for all $N$ large enough, and these
term vanish though the whole determinant does not.\\

{\bf A.2.} Looking at the determinant in (3.3.3) one can recognize primitive
characters of the group $GL(n)$ or $U(n)$ where $$t_i=\sum_k{1\over i}\e_k^i
\eqno{(A.2.1)}$$ and $\e_k$ are the eigenvalues of a matrix for which this 
character is evaluated (see [7]); it is supposed that $m_0\geq m_1\geq...\geq m_{N-1}
$. The latter can always be achieved by a permutation and relabeling of 
indices. 

It is not easy to understand why $\tau$-functions happen to be
related to characters (a good explanation of this fact can lead to new
profound theories). However, we can extract lessons from this relationship.
First of all, the $\tau$-function is given as an expansion in a series in 
characters. Hence it can be considered as a function on the unitary or the
general linear group invariant with respect to the conjugation. In the end
we will have an explicit formula giving this function which is the 
Kontsevich integral.

Secondly, the benefit of the usage of variables $\e_k$ instead of $t_i$ is 
obvious: the elaborated techniques of the theory of characters can be applied 
to the $\tau$-function as well. We shall use also an inverse matrix with the
eigenvalues $\l_i=\e_i^{-1}$. Thus, we have $$t_i=\sum_k{1\over i}\l_k^{-i}.
\eqno{(A.2.2)}$$ This change of variables is called the {\em Miwa
transformation}.

It is easy to show that $$p_l(t)=\sum_{l_1+...+l_N=l}
\e_1^{l_1}\e_2^{l_2}...\e_N^{l_N}.$$ This is the Newton formula expressing
complete symmetric functions of variables $\e_k$ in terms of sums of
their powers, $t_i$.

Introduce a notation $$|g_0(\l),...,g_{N-1}(\l)|=\det(g_j(\l_i)).$$
Then it can be proven that
$$\tau_N={|f_0(\l),f_1(\l)\l,f_2(\l)\l^2,...,f_{N-1}(\l)\l^{N-1}|\over|1,\l,
\l^2,...,\l^{N-1}|}\eqno{(A.2.3)}$$ where $f_i(\l)=\sum_0^\infty c_m^{(i)}
\l^{-m}$ being $f_i=f_{i+n}$.\\

{\bf A.3.} Up to this point there were no restrictions imposed on the
coefficients $c_m^{(k)}$ except the periodicity (A.1.1). Now we try to satisfy
the string equation (see, e.g., [9] or [10-11]). The string equation is
closely related to the so-called additional symmetries of the KP hierarchy
(each equation of the hierarchy provides a symmetry for all other equations,
they are the main symmetries, while an additional symmetry is not contained
in the hierarchy itself). The string equation is equivalent to the fact that 
$\tau$ does not
depend on the additional variable $t_{-n+1,1}^*$. However, it is known that the
derivative $\p_{-n+1,1}^*$ is defined not uniquely: there is still a
possibility for gauge transformations and for a shift of variables $t_i
\mp t_i+a_i$.

First of all, the operator $\p_{-n+1,1}^*$ acts on $\tau$, as the
operator $$ W_{-n}^{(2)}=\sum_{i+j=n}ijt_it_j-2\sum_1^\infty t_{i+n}\p_i+
(n-1)nt_n+C.$$ This
has to be expressed in terms of new variables, $\l$'s or $\e$'s. The last two
terms are not important: an arbitrary linear term in $t_i$ and a constant can
be added, this is precisely a gauge transformation. The result is
$$W_{-n}^{(2)}=\sum_{i,j>0;i+j=n}\sum_r{1\over 
\l_r^i}\sum_s{1\over\l_s^j}-2\sum_k{1\over\l_k^{n-1}}{\p\over\p\l_k}.$$
If a possibility of a shift $t_i\mp t_i+a_i$ is taking into account,
then $$\tau_N={(|f_0(\l),f_1(\l)\l,f_2(\l)\l^2,...,f_{N-1}(\l)\l^{N-1}|
\prod_i\exp\sum_j a_j\l_i^j)_{\leq N-1}\over|1,\l,\l^2,...,\l^{N-1}|}\eqno
{(A.3.1)}$$ where the subscript $\leq N-1$ means taking powers of $\l_i$ not 
larger than $N-1$. 

Skipping all calculations, the string equation implies the following
recurrent system of equations for $f_k$:
$$\l^kf_k=D^kf_0,~~k=0,1,...,n-1\eqno{(A.3.2)}$$ where
$$D=\sum_ja_jj\l^{j-n}-{n-1\over 2}{1\over\l^n}+{1\over\l^{n-1}}{\p\over\p\l}
.$$ If we recall that $f_0=f_n$ then for the first term we have
$$D^nf_0=\l^nf_0 \eqno{(A.3.3)}$$ or
$$\left(\sum_ja_jj\l^{j-n}-{n-1\over 2}{1\over\l^n}+{1\over\l^{n-1}}{\p\over\p\l}
\right)^nf_0=\l^nf_0$$ 
 which must hold identically in $\l$ can be satisfied by some
$f_0=\sum_0^\infty c_m^{(0)}\l^{-m}$ if and only if $\sum_ja_jj\l^{j-n}=\l$, 
i.e., $\sum_ja_j\l^j=\l^{n+1}/(n+1)$.

It is possible to perform some scaling transformation $\l\mp a\l,~D\mp a^{-1}D
$. This is not so important, but just in order that our formulas coincide with 
those in [5] we take $a=n^{1/(n+1)}$. Then $$D=\l+{n-1
\over 2n\l^n}+{1\over n\l^{n-1}}{\p\over\p\l}$$ $$=\l^{(n-1)/2}\exp\(-{n\over
n+1}\l^{n+1}\){\p\over\p\l^n}\exp\({n\over n+1}\l^{n+1}\)
\l^{-(n-1)/2}. \eqno{(A.3.4)}$$ The function $g_0=\exp(n\l^{n+1}/(n+1))
\l^{-(n-1)/2}f_0$ satisfies the equation $$\({\p\over\p\l^n}\)^ng_0=
\l^ng_0; \eqno{(A.3.5)}$$ which, being written in the variable $\mu=\l^n$, is a
generalization of the Airy equation (the latter is a special case with $n=2$).
A solution can be found by the Laplace method. The solution is
$$ f_0=\l^{(n-1)/2}\exp\(-{n\over n+1}\l^{n+1}\)\int\exp\(\l^nm-{m^{n+1}\over
n+1}\)dm.$$ It is easy to see that $$D^kf_0=\l^{(n-1)/2}\exp\(-{n\over n+1}\l^
{n+1}\)\int m^k\exp\(\l^nm-{m^{n+1}\over n+1}\)dm.$$ The $\tau$-function is
$$\tau_N=i^{N(N+1)/2}\prod_k\l_k
^{(n-1)/2}\exp\(-\sum_k{n\over n+1}\l_k^{n+1}\)$$ $$\times \int...
\int dm_1...dm_N\prod_{r>s}{m_r-m_s\over\l_r-\l_s}\exp\sum_k\(i\l_k^nm_k-
{(im_k)^{n+1}\over n+1}\).\eqno{(A.3.6)}$$  We do not discuss here the problems of
the choice of contours of integration and of the convergence.\\

{\bf A.4.} In a sense, the problem is already solved, the solution is explicitly
written. However, we remember the above remark that it is natural to consider a 
$\tau$-function as a function on a matrix group invariant under conjugation, 
$\l_k$ being eigenvalues of a matrix, since originally it was written as a 
series in characters of the group. If we want to restrict ourselves to real
values of the time variables $t_i$, i.e., real $\l_k$, then the function is
restricted to matrices with real eigenvalues, e.g., Hermitian. Our intention
now is to write the formula (A.3.6) more directly in terms of matrices themselves,
not of their eigenvalues. 

The theory of representations has a tool for that, the Harish-Chandra formula
([8]). Let $\Phi(X)$ be a function of Hermitian matrices $X$, invariant under
conjugations, i.e., depending only on eigenvalues of $X$. We consider integrals
over the space of Hermitian matrices (h.m.) $\int_{h.m.}\Phi(X)\exp(-i{\rm tr}~
XY)dX$ where $Y$ is a Hermitian matrix with eigenvalues $\mu_k$, and
$dX=\prod_{i,j}dX_{ij}$. If $X=UMU^{-1}$
where $U$ is unitary and $M=$diag $m_1,...,m_N$ then the function can be partly
integrated, with respect to the ``angle" variables $U$ and only integral over 
diagonal matrices $M$ remains: $$\int_{h.m.}\Phi(X)\exp(i{\rm tr}~XY)dX$$ $$=
(2\pi i)^{N(N-1)}\int...\int dm_1...dm_N\prod_{r>s}{m_r-m_s\over\mu_r-\mu_s}
\Phi(M)\exp(i\sum m_k\mu_k).$$ Using this techniques, one can show that
$$\tau_N={\rm const}{\int\exp(\tr(-{\rm n.l.}(Z+\Lm)^{n+1}/(n+1)))dZ\over
\int\exp(\tr(-{\rm quad.}(Z+\Lm)^{n+1}/(n+1)))dZ}\eqno{(A.3.7)}$$
where ${\rm n.l.}(Z+\Lm)^{n+1}$ symbolizes all terms of degree higher
than 1 in the expansion of $(Z+\Lm)^{n+1}$ in powers of $Z$ while ${\rm quad.}(Z+\Lm)^{n+1}$
stands for the quadratic term; $\Lm$ is a matrix with eigenvalues
$\l_k$.

The expression (A.3.7) is called the {\em Kontsevich integral} (more precisely, 
its generalization from $n=2$ to any $n$).  \\

{\bf References.}\\

\noindent {\bf 1.} Ueno, K. and Takasaki, K., Toda lattice hierarchy, Adv. Studies
in Pure Math. 4, 1-95, 1984\\

\noindent {\bf 2.}  Adler, M. and van Moerbecke, P., Vertex operator solutions to the
discrete KP-hierarchy, Comm. Math. Phys., 203, 185-210, 1999\\

\noindent {\bf 3.} Dickey, L. A., Modified KP and Discrete KP, Lett. Math. Phys.,
48, 277-289, 1999 \\

\noindent {\bf 4.} Kontsevich, M., Intersection theory on the moduli space
of curves and the matrix Airy function, Comm. Math. Phys., 147, 1-23, 1992\\

\noindent {\bf 5.} C. Itzykson and J.-B. Zuber, Combinatorics of the modular
group II, The Kontsevich integral, Int. J. Mod. Phys., 7(29), 5661-5705, 1992\\

\noindent {\bf 6.} M. Adler and P. van Moerbeke, A matrix solution to the
two-dimensional $W_p-$gravity., Comm. Math. Phys., 147, 25-26, 1992\\

\noindent {\bf 7.} H. Weyl, The classical groups, Princeton Univ. Press, 1939\\

\noindent {\bf 8.} Harish-Chandra, Differential operators on a semisimple Lie
algebra, Amer. J. Math., 79, 87-120, 1957\\

\noindent {\bf 9.} van Moerbeke, P., Integrable foundations of the string theory, in
{\em Lectures on Integrable systems}, ed. Babelon, Cartier, 
Kosmann-Schwarzbach, World Scientific, 1994\\

\noindent {\bf 10.} Dickey, L. A., Additional symmetries of KP, Grassmannian, and the
string equation, Modern Physics Letters A, 8, No 13, 1259-1272, 1993\\

\noindent {\bf 11.} Dickey, L. A.,
Lectures on W-Algebras, Acta Applicandae Mathematicae, 47, 243-321, 1997\\

\end{document}